\documentclass[12pt,preprint]{revtex4}
\usepackage{graphicx} 
\usepackage{amsmath}
\usepackage{amsfonts,amssymb}
\usepackage[matrix,arrow,curve,frame,poly,arc]{xy}
\usepackage[english]{babel}

\def\be{\begin{equation}}
\def\ee{\end{equation}}
\def\ba{\begin{eqnarray}}
\def\ea{\end{eqnarray}}

\begin{document}
\bibliographystyle{plainnat}

\title{The 1D parabolic-parabolic Patlak-Keller-Segel model of chemotaxis: the particular integrable case and soliton solution}

\author{Maria Shubina }


\affiliation{Skobeltsyn Institute of Nuclear Physics\\ Lomonosov Moscow State University\\ Leninskie gory, GSP-1, Moscow 119991, Russian Federation}


\begin{abstract}
In this paper we investigate the one-dimensional parabolic-parabolic Patlak-Keller-Segel model of chemotaxis. For the case when the diffusion coefficient of chemical substance is equal to two, in terms of travelling wave variables the reduced system appears integrable and allows the analytical solution. We obtain the exact soliton solutions, one of which is exactly the one-soliton solution of the Korteweg-de Vries equation.

\end{abstract}


\maketitle

\section{Introduction}
\label{intro}

In this article we consider the mathematical model of chemotaxis introduced by Patlak in the 1950s \cite{P} and by Keller and Segel in the 1970s \cite{KS1}-\cite{KS3}. Chemotaxis, or the directed movement of cells (bacteria or other organisms) along the gradient of concentration of chemical irritants, attracts a great interest. At present there is a number of very interesting results concerning the existence and properties of both regular and blow-up solutions of the above system in different dimensions \cite{PHCh}--\cite{BCD}. The simplified model has the form:
\be \left\{
\begin{aligned}
u_{t}-\nabla (\delta_{1} \nabla u - \eta_{1} u \nabla v) & =0 \\ 
v_{t}-\delta_{2} \nabla^{2} v - \eta_{2} u & =0,\\ 
\end{aligned}
\right.
\ee
where $ u(t,x) $ is the cells density, $ v(t,x) $ is the concentration of the chemical substance. The positive constants $ \delta_{1} $, $ \delta_{2} $, $ \eta_{1} $, $ \eta_{2} $ are the cells and chemical substance diffusion coefficients, the chemotaxis constant and the rate of cells chemical production respectively. After the replacement $ t \rightarrow  \delta_{1} t$ and $ v \rightarrow \frac{\eta_{1}}{\delta_{1}} v $ the coefficients in the first equation in (1) become equal to unity.

\section{Exact Solution}
\label{sec:1}

We investigate the following one-dimensional parabolic-parabolic model:
\be \left\{
\begin{aligned}
u_{t}-u_{xx}+(u v_{x})_{x} & =0 \\ 
v_{t}-\alpha v_{xx}-\beta u & =0,\\
\end{aligned}
\right. 
\ee
where $ x \in \Re, t \geq 0 $, $ u=u(x,t) $, $ v=v(x,t) $, $ \alpha, \beta $ are positive constants. This system is invariant under the one-parametric group of translation $ (x, t, u, v)\rightarrow (x+c\epsilon, t+\epsilon, u, v ) $. Rewriting (2) in terms of travelling wave variables $ y=x-ct $ we obtain the system of ordinary differential equations (ODEs); one can see, that the first equation may be integrated, and on can put $ \beta = 1 $.
The reduced system has the form:
\be \left\{
\begin{aligned}
u_{y}+cu-u v_{y}+\lambda & =0 \\ 
\alpha v_{yy}+cv_{y}+u & =0,\\
\end{aligned}
\right. 
\ee
where $ u=u(y) $, $ v=v(y) $, $ \lambda = const $.
Using the Kovalevskaya-Gambier method (see \cite{Kudryashov}), improved in \cite{ARS1}, \cite{ARS2}, we can see, that this system possess the Painlev\'e property only if $ \alpha = 2 $. Let us focus on this case. 

It is convenient to solve Eqs.(3) in terms of variable
\be
z= \frac{\kappa}{|c|}\,\,e^{-\frac{cy}{2}},
\ee
where $  \kappa >0 $ is an arbitrary constant.
Then for $ v $ and $ u $ we obtain the solutions in the form:
\ba
v=-\ln \left[  \frac{|c|}{\kappa}\,z\,Z_{\nu}^{2}(z) \right] \\ \nonumber
u=c^{2} z^{2}\left( 1-\frac{1}{4}(v_{z})^{2}\right)  -\frac{\lambda}{c}, & \text{where $ \nu^{2}=\dfrac{1}{4}-\dfrac{\lambda}{c^{3}} $.} \\ \nonumber
\ea
The function $ Z_{\nu}(z) $ satisfies the modified Bessel's equation and can be present as a linear combination of Infeld's and Macdonald's functions. 

Using the series expansion of the Bessel's functions, as well as theirs asymptotic behaviour \cite{NU}, one may obtain the following asymptotic forms for $ e^{v_{\nu}(z)} $ and $ u_{\nu}(z) $:
\ba
z \rightarrow \infty: \,\,\,\,\,\,\,
e^{v_{\nu}(z)} \rightarrow 0;\,\,\,\,
u_{\nu}(z) \rightarrow 0.
\ea

\ba
z \rightarrow 0: \,\,\,\,\,\,\,
e^{v_{\nu}(z)} \rightarrow
\begin{cases}
\infty,&\text{\,\,\, $0\leq \nu < \dfrac{1}{2}$;}\\
\dfrac{\kappa}{|c|\,C^{2}}\,\dfrac{8\pi}{(\pi+2)^2},&\text{\,\,\, $\nu = \dfrac{1}{2}$;}\\
0,&\text{\,\,\,$\nu > \dfrac{1}{2}$;}
\end{cases}
\ea
\ba
u_{\nu}(z) \rightarrow c^2\left( \nu - \frac{1}{2}\right) ,
\ea
where the expression for $ \nu = \dfrac{1}{2} $ agrees with the Eqs.(9) below.

Consider now the class of solutions with half-integer index $ \nu = n+\dfrac{1}{2} $, when $Z_{\nu}(z) $ can be expressed in hyperbolic functions. The requirement of absence of divergence $ u \rightarrow -\infty $ for finite $ z $ leads to the following form for $ Z_{n+\frac{1}{2}}(z) $:
\ba
Z_{n+\frac{1}{2}}(z)=
\begin{cases}
C z^{n+\frac{1}{2}}\left( \dfrac{d}{zdz}\right) ^{n}\dfrac{\cosh(z+\zeta) }{z}, & n=2p,    \\
C z^{n+\frac{1}{2}}\left( \dfrac{d}{zdz}\right) ^{n}\dfrac{\sinh(z+\zeta) }{z}, & n=2p+1;\,\,\, p=0,1...;
\\
\zeta=\dfrac{1}{2}\ln \dfrac{2}{\pi}, \,\,\,C=const.
\end{cases}
\ea

At first let us to consider the solutions obtained for $ e^{v_{n+\frac{1}{2} }} $ and $ u_{n+\frac{1}{2} } $ as functions of $ z $. We begin with $ n=0 $, or $ \nu = \dfrac{1}{2} $. It is interesting to present the expressions for $ e^{v_{\frac{1}{2} }(z)} $ and $ u_{\frac{1}{2} } $:
\ba
e^{v_{\frac{1}{2}}(z)} = \frac{\kappa}{C^{2}|c|}\,sech^{2}(z+\zeta) \\
u_{\frac{1}{2}}(z) = z^{2}c^{2}\,sech^{2}(z+\zeta),
\ea
where (10) appears the one-soliton solution exactly the same as the well-known one of the Korteweg-de Vries equation. The expressions for $ n \geq 1 $ become more complicated, and one can see the solitonic behaviour of $ e^{v_{n+\frac{1}{2} }(z)} $ and the curves for $ u_{n+\frac{1}{2} }(z) $ in Fig.1-Fig.2.

\begin{figure}[h!]
\begin{minipage}{0.49\linewidth}
\center{\includegraphics[width=0.75\linewidth]{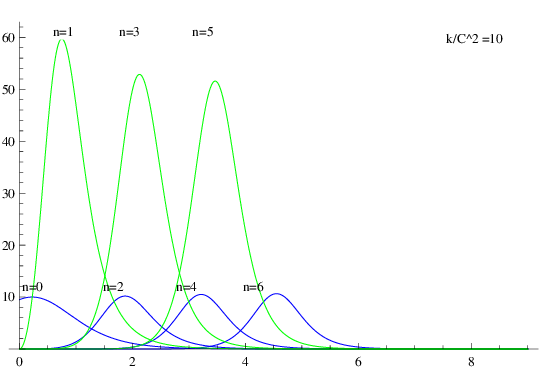} \\ Fig.1:  $ e^{v_{n+\frac{1}{2} }(z)} $; $n=0,...6$; $c=1$;}
\end{minipage}
\hfill
\begin{minipage}{0.49\linewidth}
\center{\includegraphics[width=0.75\linewidth]{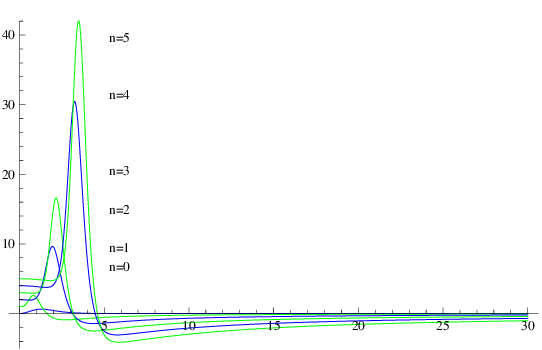} \\ Fig.2:  $ u_{n+\frac{1}{2} }(z) $; $n=0,...5$; $c=1$;}
\end{minipage}
\end{figure}

Return now to the required variable $ y $. We obtain the explicit form
of our solution by direct substitution (4) into (9), where $ \dfrac{\lambda}{c} = -c^2 n(n+1) $. The resulting formulae are complicated and slightly difficult for analytic analysis; it seems to be more convenient to present the plots. 

For $ n=0 $ in the function $ e^{v_{\frac{1}{2}}(y)} $ we have the "step" whose altitude depends on the values of velocity $ c $ and arbitrary constant $ \kappa $.
\begin{figure}[h!]
\begin{minipage}{0.49\linewidth}
\center{\includegraphics[width=0.75\linewidth]{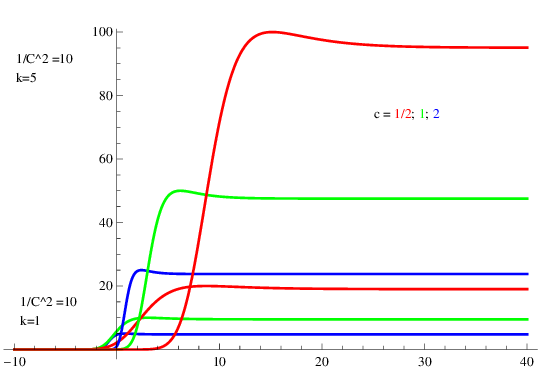} \\ Fig.3:  $ e^{v_{n+\frac{1}{2} }(y)} $; $n=0$. 
}
\end{minipage}
\hfill
\begin{minipage}{0.49\linewidth}
\center{\includegraphics[width=0.75\linewidth]{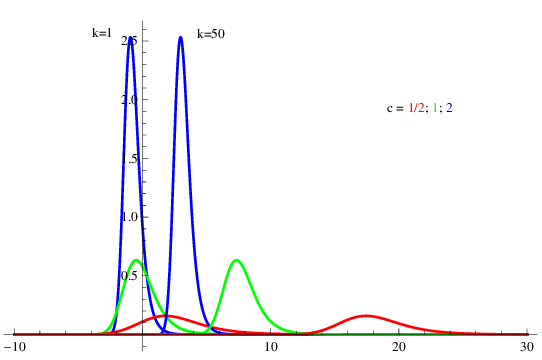} \\ Fig.4:  $ 
u_{n+\frac{1}{2} }(y) $; $n=0$. 
}
\end{minipage}
\end{figure}
One may see that these curves become higher and shift to the right with different rates for rising $ \kappa $. The $ u_{\frac{1}{2}}(y) $ is the positive function whose altitude and sharpness of peak depends on $ c $  (see Fig.3 - Fig.4).

For $ n \geq 1 $ we can see that the solitonic behaviour of $ e^{v_{n+\frac{1}{2} }(y)} $ is retained for different values of $ c $ and $ \kappa $; the curves  become higher, more tight and they shift to the right also with increase of $ c $ and $ \kappa $. For the cells density $ u_{n+\frac{1}{2} }(y) $ the obtained solution has the negative  section converging to zero for $ cy \rightarrow -\infty $ (Figs.5--8). 

\begin{figure}[h!]
\begin{minipage}{0.49\linewidth}
\center{\includegraphics[width=0.75\linewidth]{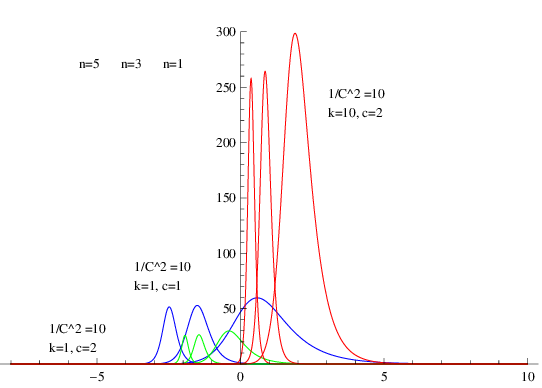} \\ Fig.5:  $ e^{v_{n+\frac{1}{2} }(y)} $; $n=1;3;5$. 
}
\end{minipage}
\hfill
\begin{minipage}{0.49\linewidth}
\center{\includegraphics[width=0.75\linewidth]{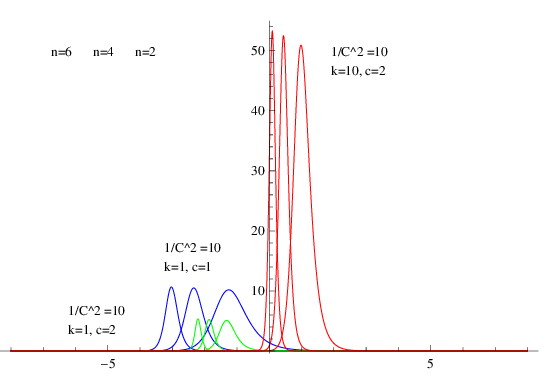} \\ Fig.6:  $ e^{v_{n+\frac{1}{2} }(y)} $; $ n=2;4;6 $. 
}
\end{minipage}
\end{figure}

\begin{figure}[h!]
\begin{minipage}{0.49\linewidth}
\center{\includegraphics[width=0.75\linewidth]{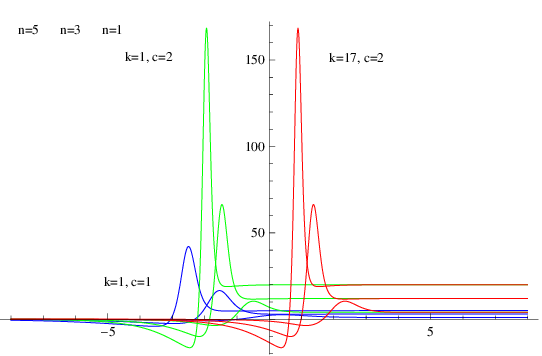} \\ Fig.7:  $ u_{n+\frac{1}{2} }(y) $; $n=1;3;5$. 
}
\end{minipage}
\hfill
\begin{minipage}{0.49\linewidth}
\center{\includegraphics[width=0.75\linewidth]{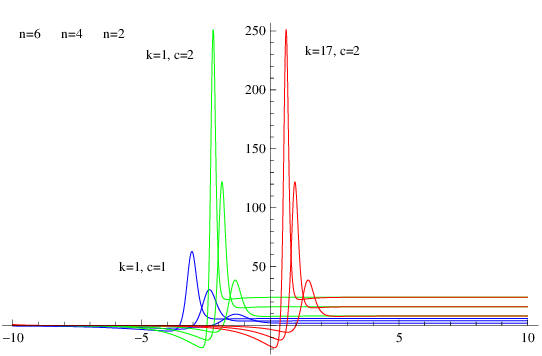} \\ Fig.8:  $ u_{n+\frac{1}{2} }(y) $; $ n=2;4;6 $. 
}
\end{minipage}
\end{figure}

The curves for the concentration of the chemical substance $ v_{n+\frac{1}{2}}(y) $ are presented in the Fig.9. Since $ v_{n+\frac{1}{2} }(y) $ have to be positive (nonnegative) we see, that these functions do not satisfy this requirement in all domain of definition. 

\begin{figure}[h!]
\begin{minipage}{0.49\linewidth}
\center{\includegraphics[width=0.75\linewidth]{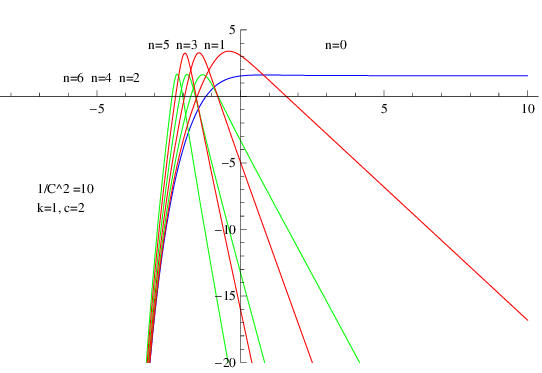} \\ Fig.9:  $ v_{n+\frac{1}{2} }(y) $; $n=0, 1, ... 6 $.
}
\end{minipage}
\hfill
\begin{minipage}{0.49\linewidth}
\center{\includegraphics[width=0.75\linewidth]{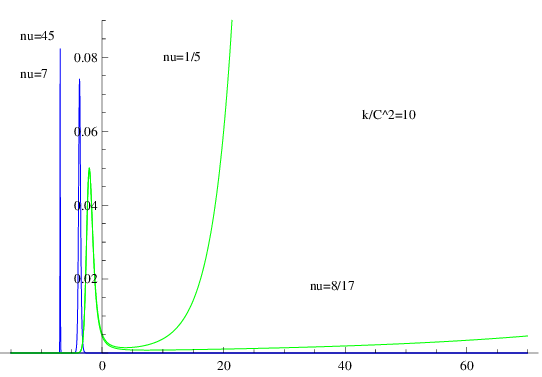} \\ Fig.10:  $ e^{v_{\nu}(y)} $; $\nu=1/5; 8/17; 7; 45$.
}
\end{minipage}
\end{figure}

\begin{figure}[h!]
\begin{minipage}{0.49\linewidth}
\center{\includegraphics[width=0.75\linewidth]{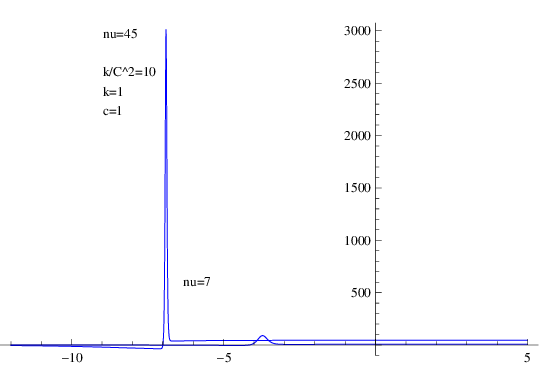} \\ Fig.11:  $ u_{\nu}(y)$; $\nu=7; 45 $.

}
\end{minipage}
\hfill
\begin{minipage}{0.49\linewidth}
\center{\includegraphics[width=0.75\linewidth]{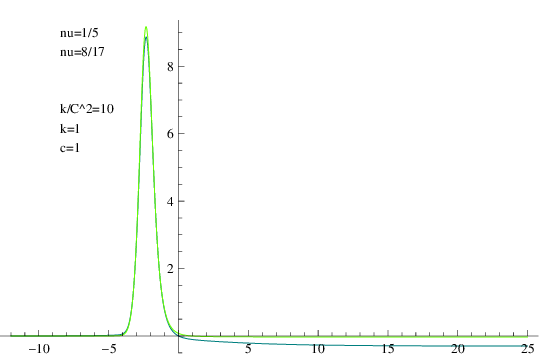} \\ Fig.12:  $ u_{\nu}(y) $; $\nu=1/5; 8/17$.
}
\end{minipage}
\end{figure}

In conclusion it seems to be interesting to present the plots for $ e^{v_{\nu}(y)} $ and $ u_{\nu} (y) $ for different values of $ \nu $, Fig.10-Fig.12. It is interesting to see, that there are irregular solutions for $ e^{v_{\nu}(y)} $, however the corresponding solutions for $ u_{\nu}(y) $ are regular (see (6)-(8)).

\section{Conclusion}
\label{sec:2}

We investigate the one-dimensional parabolic-parabolic Patlak-Keller-Segel model. One of the reductions of this system to ODEs turns out to be integrable. This corresponds to the case when the chemical substance diffusion coefficient is equal to $ 2 $. After integration we obtain the exact soliton solutions in terms of travelling wave variables. The interesting fact is that in the plane $ (y, e^{v(y)}) $ one of the solution above coincides with the well-known Korteweg-de Vries one. However it is very likely that the above solutions are model only and they are not useful in practice, and it is interesting to find more useful solutions. Further it would be interesting to analyze in more depth the general mathematical properties of this system. All these questions require a further investigations.

\end{document}